# Dislocation Activities at the Martensite Phase Transformation Interface in Metastable Austenitic Stainless Steel: An *In-situ* TEM Study

*Jiabin Liu[2], Qiong Feng[3], Xiaoyang Fang[2], Hongtao Wang[1*], Jian Lu[4], Dierk Raabe[5*], Wei Yang[1]*

1 Institute of Applied Mechanics, Zhejiang University, Hangzhou 310027, China

2 College of Materials Science and Engineering, Zhejiang University, Hangzhou 310027, China

3 China Railway Electrification Survey Design & Research Institute Co. Ltd, Tianjing 300250, China

4 Department of Mechanical and Biomedical Engineering, City University of Hong Kong, Hong Kong, China

5 Max-Planck-Institut für Eisenforschung GmbH Max-Planck-Str. 1, 40237 Düsseldorf, Germany

---

[*] Correspondence and requests for materials should be addressed to H. Wang (email: htw@zju.edu.cn), or D. Raabe (email: d.raabe@mpie.de).

**Abstract**

Understanding the mechanism of martensitic transformation is of great importance in developing advanced high strength steels, especially TRansformation-Induced Plasticity (TRIP) steels. The TRIP effect leads to enhanced work-hardening rate, postponed onset of necking and excellent formability. *In-situ* transmission electron microscopy has been performed to systematically investigate the dynamic interactions between dislocations and α′ martensite at microscale. Local stress concentrations, *e.g.* from notches or dislocation pile-ups, render free edges and grain boundaries favorable nucleation sites for α′ martensite. Its growth leads to partial dislocation emission on two independent slip planes from the hetero-interface when the austenite matrix is initially free of dislocations. The kinematic analysis reveals that activating slip systems on two independent {111} planes of austenite are necessary in accommodating the interfacial mismatch strain. Full dislocation emission is generally observed inside of austenite regions that contain high density of dislocations. In both situations, phase boundary propagation generates large amounts of dislocations entering into the matrix, which renders the total deformation compatible and provide substantial strain hardening of the host phase. These moving dislocation sources enable plastic relaxation and prevent local damage accumulation by intense slipping on the softer side of the interfacial region. Thus, finely dispersed martensite distribution renders plastic deformation more uniform throughout the austenitic matrix, which explains the exceptional combination of strength and ductility of TRIP steels.

# 1. Introduction

Among various deformation mechanisms, the martensitic transformation of austenite under mechanical loading has been known for a long time, inspiring the development of TRansformation-Induced Plasticity (TRIP) steels [1-4]. The TRIP effect leads to enhanced work-hardening rate, postponed onset of necking and thus excellent formability. It has attracted high interest in the underlying thermodynamics and atomic mechanisms associated with the phase transformation from austenite to $\alpha'$ martensite, such as for instance outlined in the Bogers-Burgers-Olson-Cohen (BBOC) models [5, 6], which have greatly advanced our basic understanding in this field.

Following the high demand for weight reduction in the automotive industry, better understanding of the strain hardening mechanisms in such advanced high strength steels has become a paramount requirement. For achieving improved combinations of strength and ductility, the role of the microstructure, especially of dislocation multiplication and interaction at the austenite - $\alpha'$ martensite interface, must be better understood. For this purpose, it is required to not only study the phase transformation itself but also the complex interactions emerging from the dynamically evolving hetero-phase microstructures in-situ during deformation, which account for the high strain hardening capacity that is associated with the TRIP effect.

Among the various microstructures, dislocations play a key role in TRIP steels[7-9], carrying the major portion of the plastic strain. The volume of the $\alpha'$ martensite is larger (about 4%) than that of the austenite from which it forms[10]. The shape change and the volume increase must be accommodated by the generation and motion of dislocations in the surrounding austenite[11]. A high density of pre-stored dislocations in the austenite may pin the phase boundary by obstructing the required cooperative movement of atoms during the $\alpha'$ martensitic transformation, known as mechanical stabilization [12, 13]. An opposite effect may also be expected, namely, in that the associated high back stress stemming from a high dislocation density may tend to promote the α′ martensitic transformation [14]. It is not clear up to now which of these effects prevails in TRIP steels.

To answer these questions, in this study *in-situ* transmission electron microscopy (TEM) has been carried out to systematically investigate the interaction of dislocation activities with the martensitic transformation, using metastable austenitic stainless steel as a model material. We demonstrate that the nucleation and growth behavior of the α′ martensite are closely related to the dislocation activity at the transformation interfaces, which explains the macroscopically observed pronounced strain hardening capacity of TRIP steels.

## 2. Experimental

The commercial AISI 301LN stainless steel was supplied by Outokumpu, Finland. The composition, measured by inductively coupled plasma mass spectrometry, is determined to be 6.02 wt.% Ni, 16.61 wt.% Cr, 1.76 wt.% Mn, 0.18 wt.% N, 0.14 wt.% N, 0.52 wt.% Si and balanced by Fe. The initial material was cut from 1 mm-thick sheets, solid solution treated at 1050 ºC for 2 h with Ar ambient and then quenched in water. The microstructure obtained contains equiaxed grains with an average size of 54.5 μm. *In-situ* TEM experiments were done using a straining holder (Gatan 654) equipped in a JEM-2100 TEM operated at 200 kV. The specimens were strained by controlling the total elongation *via* a step motor in the straining holder. The deformation process was recorded by a Gatan 831 CCD camera at a rate of 2 frames/s. TEM samples were prepared by a twin-jet electro-polisher using an electrolyte of 10 vol.% perchloric acid and 90 vol.% acetic acid at a temperature of 10 ºC. *Ex-situ* tensile tests were carried out using a universal testing machine (MTS Alliance RT/30) at room temperature and an initial strain rate of $1\times10^{-3}$ s$^{-1}$. The dimensions of the tensile specimens were 6.0 mm in width and 25.0 mm in gauge length. The initial thickness before rolling is 1 mm. At least three samples were tested for each state for ensuring reproducibility.

## 3. Results and discussion

### 3.1. *In-situ* TEM study on the stress induced α′ martensite

It has been well documented that α′ martensite is frequently found at intersections of two inclined ε-martensite bundles [7, 15-17]. These observations provided strong support of models that explain α′ martensite formation by a process consisting of two successive shears, *e.g.* the BBOC model, which involves a first 1/3 face centered cubic (FCC) twinning shear of austenite and an ensuing 1/2 FCC twinning shear[18, 19]. Further studies revealed that α′ martensite can nucleate in many other situations due to the complex microstructure evolution during deformation, such as intersections between ε platelets and twins or grain boundaries, or even inside a single ε platelet [17, 20]. On the other hand, direct transformation from γ to α′ has also been identified, yet, under high stress levels[21]. The incompatible phase strain leads to enriched dislocation activities at the transformation interfaces, which is closely related to the structure of α′ martensite.

Figure 1(a) shows a typical TEM sample for in-situ straining, prepared by the double-jet electro-polishing method. The electron transparent region has an annular shape, as roughly specified by the dashed borderlines. Martensite generally nucleates in front of irregular notches on the inner circumference, marked as ROI (region of interest), due to stress concentration under straining. Further deformation causes rapid growth of martensite with complex dislocation structures ahead of these moving hetero-interfaces (Fig. 1(b)). Away from the phase transformation front, the γ/α′ orientation was determined to be (111)γ // (110)α′ with [-110]γ // [1-11]α′ (Fig. 1(c)), in agreement with the Kurdjumov-Sachs orientation relationship (K-S OR). It is noted that only K-S OR was observed in all of our in-situ straining experiments, where martensites are induced by concentrated stress. This observation agrees with early investigations on isothermal martensitic transformation of Fe-Ni-Mn alloys under applied stress[20-24]. The complex structure of the phase front is revealed by bright-field TEM image in Fig. 1(d). To enhance the diffraction contrast, the electron beam is aligned to a high-index zone axis of one variant (Figs. 1(e-f)). Figure 1(d) shows intercalation between two variants, as indicated by α$_1$′ and α$_2$′. A dark banded structure was observed when aligning the beam to the common direction [01-2]α′ (Fig. 1(g)). The inset diffraction pattern shows that the two variants form a twin structure about the twin plane (-1-21)α′. The dark-field images (Figs. 1(h-i)) clearly show that neighboring bands belong to different twinned variants and the wavy boundaries are not coincident with any specific lattice plane. As revealed in Fig. 1(d), the variant thickness increases with distance from the martensite nucleus tip, developing a fine-to-coarse twin structure. We note that only one variant will be ultimately dominant during growth with some small remaining patches of fine twin stacks.

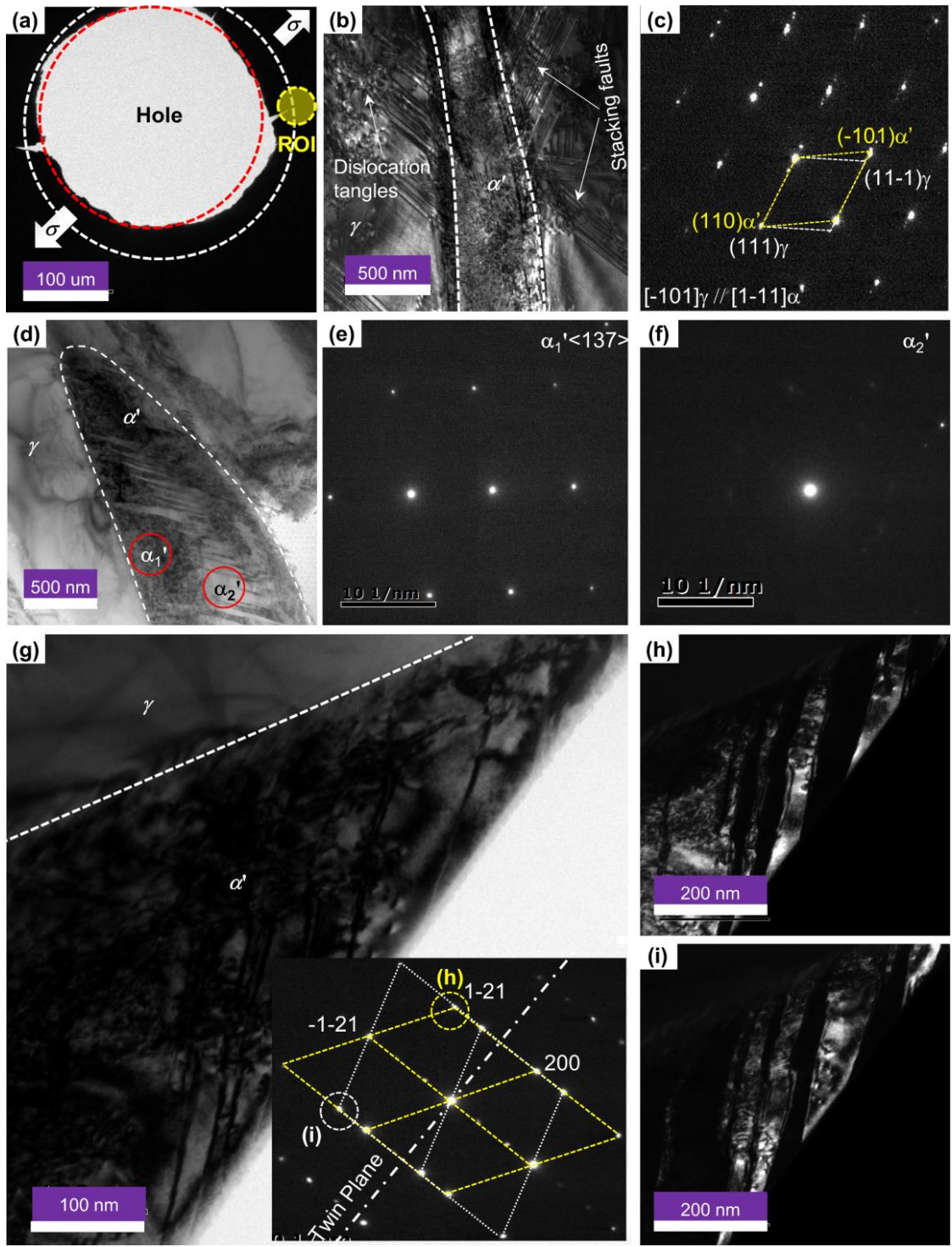

Figure 1. (a) TEM sample exposed to in-situ straining. The arrows indicate the loading direction. The region of interest (ROI) is generally located ahead of notches. (b) The α′ martensite nucleates directly from an original pristine austenite matrix. Complex dislocation structures develop during the rapid growth under straining. (c) The corresponding electron diffraction pattern from both martensite and austenite. (d) A bright-field TEM image shows that the phase front has two variants. (e) and (f) are the corresponding electron diffraction patterns from $\alpha_1'$ and $\alpha_2'$, respectively. (g) A bright-field TEM image shows that $\alpha_1'$ and $\alpha_2'$ form a twin structure.

Inset is the corresponding electron diffraction pattern. (h) and (i) are the corresponding dark-field images obtained from the reflections encircled in the inset to (g), respectively.

**3.2. Partial dislocation emission from γ/α′ interfaces**

In general, a wedge-shaped tip is nucleated ahead of notches in the pristine austenite. Further growth retains the wedge shape and activates partial dislocations from both sides on two independent slip planes (Fig. 2(a)). The martensite diffraction pattern was indexed for [011]α′ lying in the twin plane (-2-11)α′ of both variants (Fig. 2(b)). The remaining spots stem from double diffraction and have similar intensity as the major reflections, suggesting a stacking geometry of the two variants with approximately equal volume fraction. The matrix diffraction pattern was indexed for the [-233]γ zone axis (Fig. 2(c)), which is inclined to all of the four slip planes. This setup helps to visualize dislocation activities in the austenite matrix. The specific austenite/martensite orientation observed for the martensite nucleus was determined to be [-233]γ // [011]α (Fig. 2(d)), satisfying the K-S orientation relation.

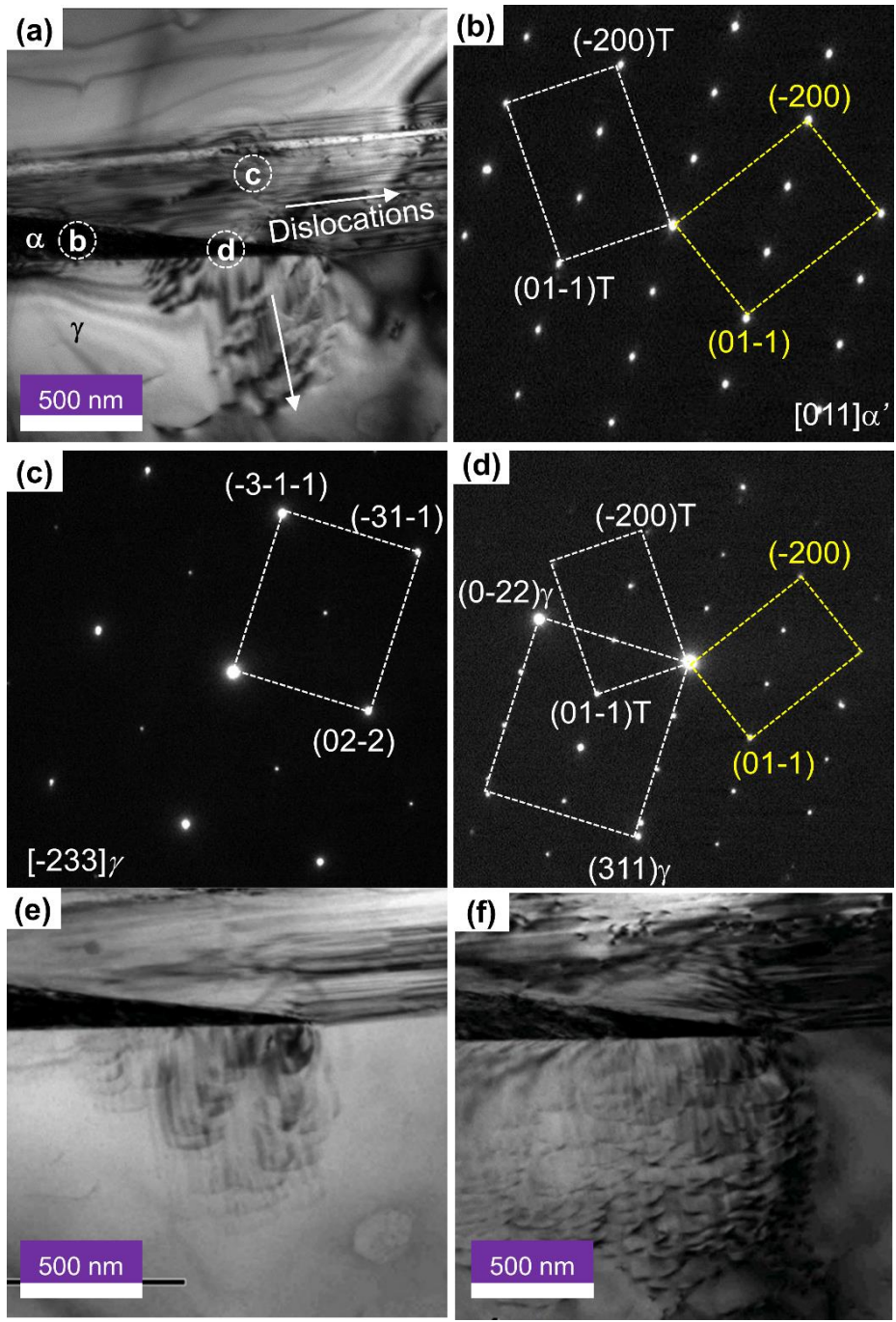

Figure 2. (a) TEM image of a martensite tip growing into the austenite. Two sets of dislocations were observed in situ gliding away from the phase interface. Their glide direction is indicated by arrows. (b) SAED patterns corresponding to the martensite tip, showing typical {211} twin diffraction patterns with <011> zone axis. (c) SAED patterns corresponding to the region (c) in (a), indicating the $[\bar{2}33]\gamma$ zone axis. (d) SAED patterns corresponding to the zone (d) in (a), showing $[\bar{2}33]\gamma$ // $[011]\alpha'$. (e) and (f) are sequential video clips showing large amount of partial dislocation emission from the hetero-interface.

The in-situ TEM reveals that two sets of partial dislocations are emitted from the moving transformation interface (Figs. 2(e-f)). This effect is a direct consequence of the incompatible phase transformation strain and it is closely related to the relative orientation between the martensite nucleus and the austenite matrix. The TEM investigation reveals both of the twinned martensite variants ($\alpha_1'$ and $\alpha_2'$) to assume K-S OR to the austenite matrix. Figure 3(a) shows a constructed atomic model according to the TEM investigation. For clarity, Figs. 3(b-d) show the two-dimensional projection of $\alpha_1'/\gamma$, $\alpha_2'/\gamma$ and $\alpha_1'/\alpha_2'$ along $[110]\gamma$ // $[111]\alpha_1'$ // $[111]\alpha_2'$ and $[012]\alpha_1'$ // $[012]\alpha_2'$, respectively. Both K-S ORs are satisfied since $(-110)\alpha_1'$ // $(1-11)\gamma$ and $(-110)\alpha_2'$ // $(-111)\gamma$. The two variants are symmetric about the plane $(1-12)\alpha'$ (Fig. 3(d)). To extract the information of the interface-emitted geometrically necessary dislocations, the phase transformation operation, which maps a FCC lattice into a BCC lattice, was cast into a simple form of sequential operations, *i.e.* a sequence of simple shears along certain slip systems. Mathematically, the phase transformation tensor (**T**) can be decomposed as **T** = $\eta$**DS**$_1$**S**$_2$, where $\eta$ accounts for the isotropic volume expansion, **D**, **S**$_1$ and **S**$_2$ the lattice distortion. **S**$_1$ and **S**$_2$ delineate the shear transformation along two independent slip systems. However, the product of **S**$_1$ and **S**$_2$ cannot transform an FCC lattice into an ideal BCC one. The deviation from the ideal BCC lattice needs to be corrected by the deformation tensor **D**. A schematic decomposition is shown in Fig. 3(e). The von Mises equivalent strain ($\varepsilon_M$) can be introduced as a scalar measure of the magnitude of the shear transformation or deformation, *i.e.*

$$\varepsilon_M = \sqrt{1/2 \mathbf{E'}:\mathbf{E'}} \qquad (1)$$

where **E**′ denotes the deviation part of the Green-Lagrange strain **E** = ½ (**F**$^T$**F** − **I**) and the deformation gradient **F** represents **D**, **S**$_1$ or **S**$_2$. As the shear transformation components **S**$_1$ and **S**$_2$ are designed to be related to the geometrically necessary dislocation content, we expect $\varepsilon_M(\mathbf{D}) \ll \varepsilon_M(\mathbf{S}_1)$ or $\varepsilon_M(\mathbf{S}_2)$. Taken the FCC lattice coordinate as the global reference coordinate system, $\mathbf{S}_1$ and $\mathbf{S}_2$ can be expressed as $\mathbf{S}_i(n) = \mathbf{R}_i^T \mathbf{S}(n) \mathbf{R}_i$ where $i = 1$ or 2, $\mathbf{R}_i$ denotes the transformation matrix from a local slip system coordinate to the global coordinate, and $\mathbf{S}(n)$ is the average shear that equates to slipping one partial dislocation per $n$ {111} planes in the local slip system coordinate, *i.e.*

$$\mathbf{S}(n) = \begin{bmatrix} 1 & 0 & s/n \\ 0 & 1 & 0 \\ 0 & 0 & 1 \end{bmatrix} \text{ with } s = \frac{1}{\sqrt{2}} \quad (2)$$

By choosing two slip systems as pair, *e.g.* $b_1$ and $b_2$ in Fig. 3(f), the product $\mathbf{S}_1(n)\mathbf{S}_2(m)$ transforms an FCC lattice into an intermediate body-centered lattice (Fig. 3(e)). Here, we note that the positive integers $m$ and $n$ are independent optimization variables. The corresponding lattice vectors are denoted as $a$, $b$ and $c$. Based on Thompson tetrahedron in Fig. 3(f), the coordinate transformation between the local slip systems $b_1$ and $b_2$ and the global reference coordinate can be directly written in the form as

$$\mathbf{R}_1 = \begin{bmatrix} \frac{1}{\sqrt{6}}\begin{pmatrix} -2 \\ -1 \\ -1 \end{pmatrix} & \frac{1}{\sqrt{2}}\begin{pmatrix} 0 \\ 1 \\ -1 \end{pmatrix} & \frac{1}{\sqrt{3}}\begin{pmatrix} -1 \\ 1 \\ 1 \end{pmatrix} \end{bmatrix}, \mathbf{R}_2 = \begin{bmatrix} \frac{1}{\sqrt{6}}\begin{pmatrix} 2 \\ 1 \\ -1 \end{pmatrix} & \frac{1}{\sqrt{2}}\begin{pmatrix} 0 \\ 1 \\ 1 \end{pmatrix} & \frac{1}{\sqrt{3}}\begin{pmatrix} 1 \\ -1 \\ 1 \end{pmatrix} \end{bmatrix} \quad (3)$$

which highlights the characteristic directions of the Burgers vectors of the partial and full dislocations, and the norm of the slip planes. The coefficients in front of the inner parentheses, *i.e.* $1/\sqrt{6}$, $1/\sqrt{2}$ and $1/\sqrt{3}$, are used for the purpose of normalization. By definition, the deformation tensor $\mathbf{D}$, mapping the intermediate body-centered lattice into an ideal BCC one with the same unit cell volume (Fig. 3(e)), can be simply derived as

$$\mathbf{D} = V^{-2/3}[\mathbf{b}\times\mathbf{c} \quad \mathbf{c}\times\mathbf{a} \quad \mathbf{a}\times\mathbf{b}] \quad (4)$$

where $V = (\mathbf{a}, \mathbf{b}, \mathbf{c})$ is the unit cell volume. Clearly, the determinant of $\mathbf{D}$ is unit. For simplicity, we take the austenite lattice constant ($a_0 = 3.58$ Å) as the length unit, which directly gives $V = ½$. The corresponding von Mises strain $\varepsilon_M(\mathbf{D})$ is a scalar measure of the magnitude of distortion of the intermediate lattice as compared to the ideal BCC counterpart. By minimizing the von Mises strain $\varepsilon_M(\mathbf{D})$, we get $m = n = 3$, $\varepsilon_M(\mathbf{D}) = 2.9\%$, the intermediate body-centered lattice constants $a = 0.7919a_0$, $b = 0.7935a_0$, $c = 0.7971a_0$, $\angle 1 = 89.62°$, $\angle 2 = 88.46°$, $\angle 3 = 87.02°$, and the intermediate BCC lattice constant $a_i = 0.7937a_0$. As the martensite phase lattice constant is 2.859Å (*i.e.* $0.7986a_0$), the volume expansion coefficient $\eta$ is determined to be 1.0062. It is noted that the von Mises strain $\varepsilon_M(\mathbf{S}_1(3)) = \varepsilon_M(\mathbf{S}_2(3))$ and $\varepsilon_M(\mathbf{S}_1(3)\mathbf{S}_2(3))$ are 11.9% and 18.1%, respectively. To form the twin martensite variant, the other pair of slip systems are uniquely determined, *i.e.* $c_1$ and $c_2$ in Fig. 3(f), which lie in the same slip planes as $\boldsymbol{b}_1$ and $\boldsymbol{b}_2$, respectively. The associated large shear strain for both variants, especially imposed by the operations $\mathbf{S}_1$ and $\mathbf{S}_2$, needs to be accommodated by partial dislocations emitting from the hetero-interface into the parent phase. The above theoretical analysis suggests that such process can be accomplished by two pairs of partial dislocations slipping on two independent close-packed planes. This leads to the conclusion that activating slip systems on two independent {111} planes of austenite are necessary in accommodating the interfacial mismatch strain between the two twinned martensite variants and the austenite, as schematically shown in Fig. 3(g). This agrees with our TEM investigations that partial dislocation activities are generally observed on two independent {111} planes.

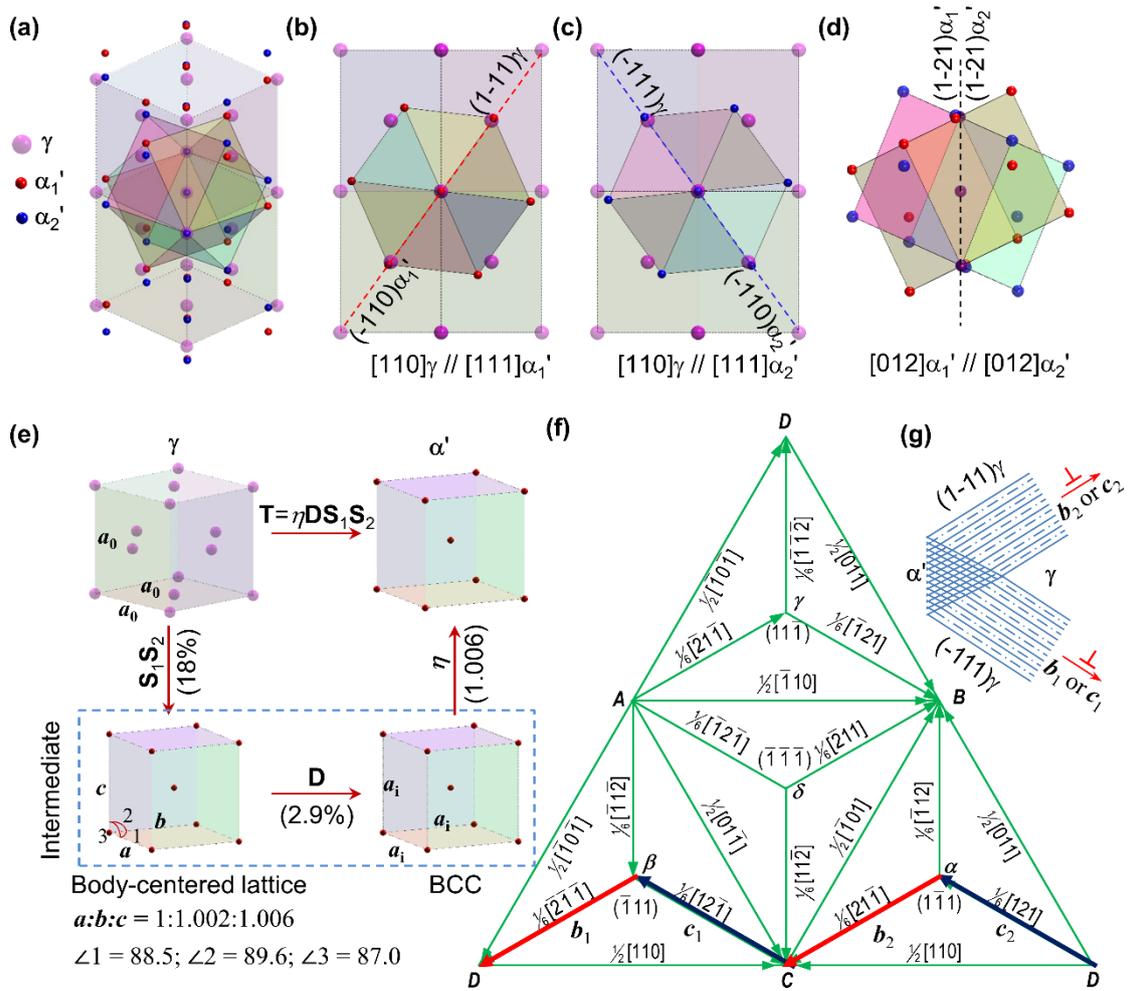

Figure 3 (a) A three-dimensional atomic model shows the orientation relations among γ, $\alpha_1'$ and $\alpha_2'$. (b) and (c) are two-dimensional projections of $\alpha_1'/\gamma$ and $\alpha_2'/\gamma$ along $[111]\alpha_1'$ and $[111]\alpha_2'$, respectively. (d) Two-dimensional projection of $\alpha_1'/\alpha_2'$ along the common direction $[012]\alpha'$. (e) A schematic decomposition of the mapping γ→α'. The shear transformation ($S_1S_2$) accounts for an equivalent von Mises strain of 18%, which is much larger than the equivalent von Mises strain of 2.9% for the deviation term (**D**). The coefficient $\eta$, accounting for volume expansion from the intermediate BCC lattice to α' martensite, is determined to be 1.006. (f) A Thompson tetrahedron shows the slip systems in an FCC lattice. Two pairs of slip systems, ($b_1$, $b_2$) and ($c_1$, $c_2$), are used to construct the twinned variants $\alpha_1'$ and $\alpha_2'$ with K-S ORs to the austenite matrix. (g) The schematic picture shows that emission of partial dislocations $b_1$ / $c_1$ and $b_2$ / $c_2$ can accommodate the major component $S_1$ and $S_2$ of the incompatible phase transformation strain.

### 3.3. Full dislocation emission from γ/α′ interfaces

Full dislocation emission is generally observed inside of austenite matrix regions that contain a high dislocation density. Figure 4 shows a typical process of dislocation activity ahead of an α′ martensite hetero-interface. Dislocations are continuously generated, bowed out and detached from the austenitic region close to the phase boundary. Sequential snapshots (Figs. 4 (a-c)) show the propagation of curved dislocation segments, as sketched in Fig. 4(d). Those dislocations are not emitted directly out of the hetero-interface but that they are formed from Frank-Read sources or bow out from existing dislocations that are located as segments in front of the hetero-interface due to the high misfit elastic stress during transformation. This 'dislocation pumping' process as enforced by the boundary conditions imposed by the moving α′ martensite, as outlined above, will raise the dislocation density inside the γ phase. When advancing the hetero-phase boundary, large amounts of dislocations are pumped into the surrounding matrix into which the martensite portion grows, sustaining the overall deformation and leading to substantial strain hardening of the host austenite phase.

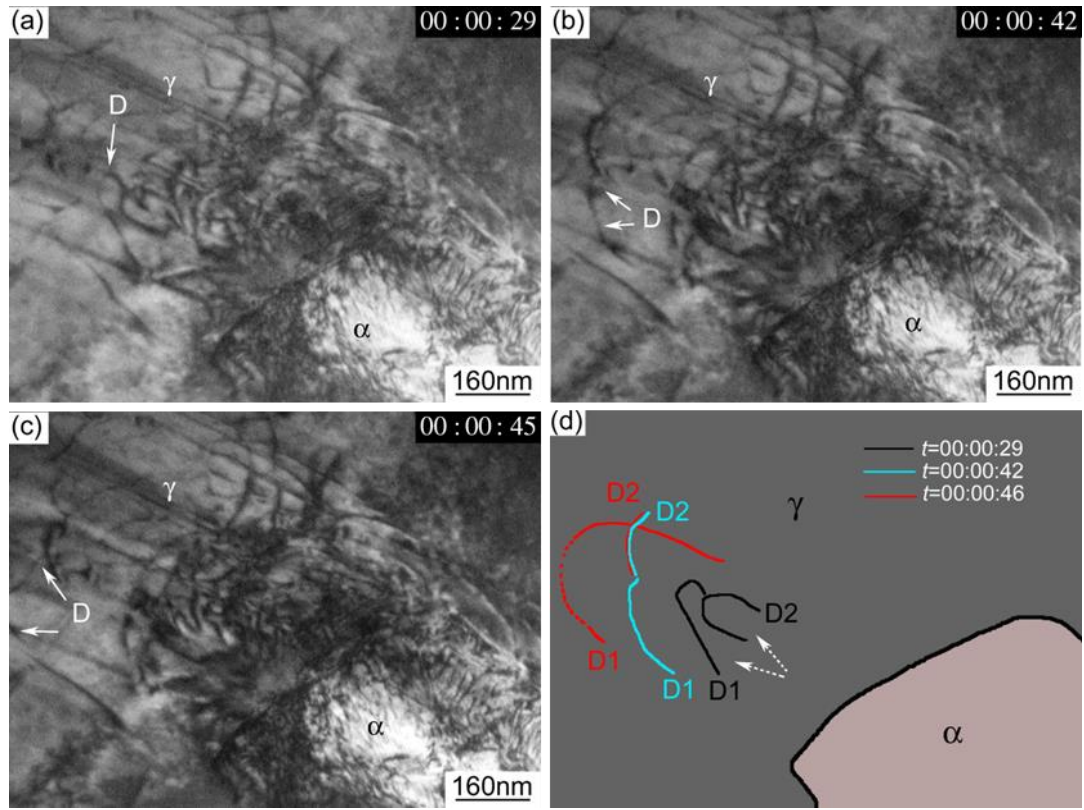

Figure 4. (a) – (c) Sequential snapshots show that large numbers of dislocation lines are pushed out in front of the austenite / martensite interface. (d) Schematic diagram showing the successive motion of the dislocation lines marked with letter D in (a) to (c). See Movie S1 for details.

It is noted that the moving FCC/BCC boundary profoundly distinguishes TRIP steels from metal composites without dynamic phase transformations. In that context, it was observed that high interfacial strength contributes to the combined ductility and strength of TRIP steels, while weak interfacial bonding accounts for reduced ductility in most metal composites [25-28]. Dislocation pile-ups before the phase boundaries due to incompatible transformation strains which need to be compensated by plastic relaxation deformation, create stress concentration spots and can cause interfacial cracks as precursors to premature failure [29, 30]. Both, high imposed local stresses and extensive dislocation glide promote the α′ martensite transformation [14, 16, 20, 31]. However, different than in the case of sessile hetero-interfaces (such as in dual phase steels) the motion of the austenite/martensite phase boundaries in TRIP steels prevents local damage accumulation stemming from intensive localized glide on the softer side of the interfacial region, *i.e.* in the austenite. Furthermore, moving dislocation sources spread over the entire austenite matrix as a direct consequence of a highly dispersed α′-martensite distribution [26]. This renders plastic deformation more uniform throughout the austenitic matrix and explains the beneficial combination of strength and ductility of TRIP steels.

### 3.4. Dislocation pile-up leads to α′ martensite nucleation

High stresses stemming from dislocations are promoting martensite nucleation are observed by our *in-situ* TEM studies. Dislocation pile-up against grain boundaries increases the local stress concentration that may activate dislocation multiplication and glide in the neighboring grain and/or facilitate martensite transformation. One example is given in Fig. 5. The data reveal that dislocation arrays glide towards a grain boundary when exposed to an *in-situ* straining situation (Fig. 5a). Continuous pile-up of the dislocation arrays activates two inclined slip systems in the adjacent grain, incubating a martensite nucleus (Figs. 5(b)-(d)). The corresponding SAED patterns are well indexed and can be attributed to the $[111]α′$ zone axis. The effect of high local stress peaks from dislocation pile-up may act twofold. From the view of energetics, higher local stress lowers the remaining required energy barrier for the onset of phase transformation[32]. On the other hand, more slip systems can be activated and the intersection of partial slips on two close-packed planes resembles the BBOC mechanism[6]. A simple manipulation by sequential shear $\mathbf{S}_1(m)$ and $\mathbf{S}_2(n)$ always transforms the FCC lattice into a distorted body-centered lattice when *m* and *n* are small positive integers (*e.g.* 1 to 5). A further relaxation will help restore the distorted intersecting region into α′ martensite. From both regards, the overall effect of a preexisting high dislocation density facilitates martensite transformation, as also revealed by the *ex-situ* experiments.

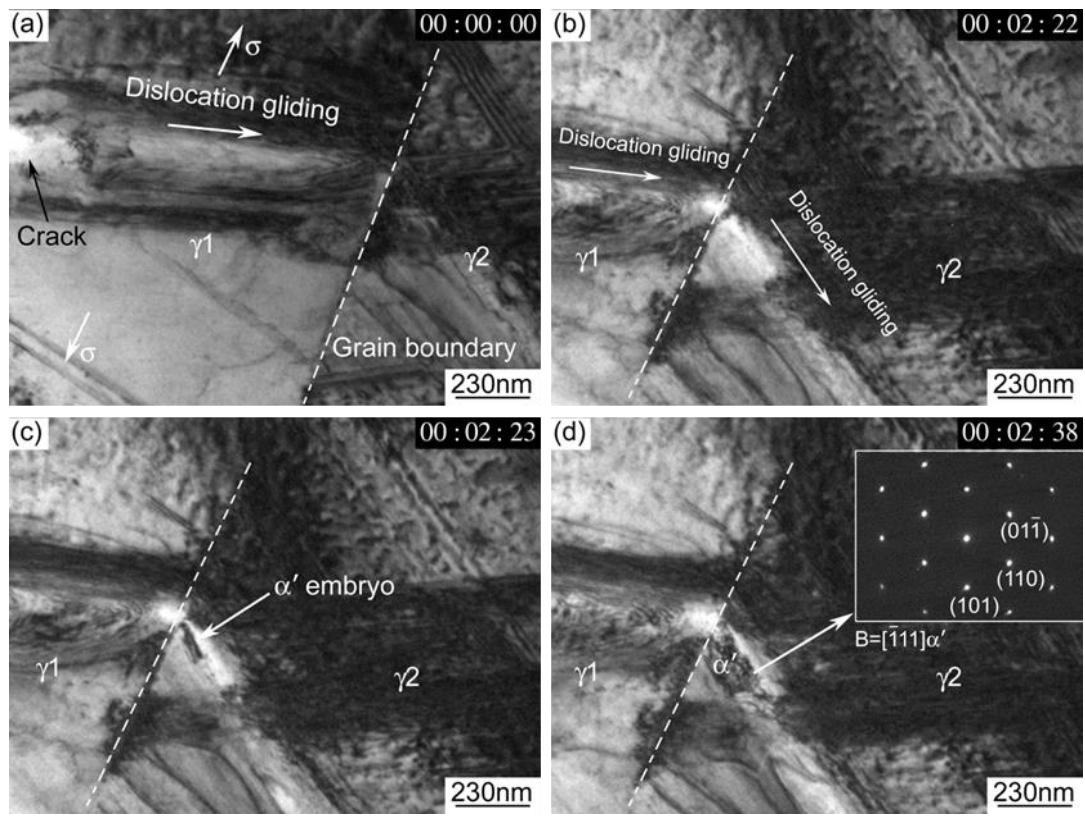

Figure 5. (a) – (c) Sequential snapshots of α′ martensite nucleation near a pre-existing grain boundary. (d) Bright field TEM image of the nucleated α′ martensite together with the corresponding SAED patterns (inset). See MovieS2 for details.

## 3.5. Overall interactions between dislocation and α′ martensite by *ex-situ* tensile experiments

In earlier works a high density of pre-existing dislocations in the austenite was assumed to increase the resistance against α′ martensite growth by obstructing the cooperative atomic displacement during phase transformation [10, 12]. It was suggested that dislocation debris, especially entangled immobile dislocations, may pin or slow down the motion of the phase boundaries. However, our current *in-situ* investigation shows that a pre-existing high dislocation density does not suppress the α′ martensite transformation. We find instead that high dislocation densities can be generated in order to accommodate the phase transformation strains by shear relaxation before the moving interface independent of the preexisting density of statistically stored dislocations inside the austenite affected. As outlined above, this effect is due to the kinematic requirement that dislocations must be emitted or pumped into the austenite matrix to compensate for the strain misfit associated with the athermal phase transformation. An *ex-situ* study has been carried out to further investigate the overall interaction between dislocations and α′ martensite. For this purpose, hot rolling was employed to inject dislocations into metastable austenitic steels while suppressing the phase transformation. The elevated hot rolling temperature raises the stacking fault energy and narrows the stacking fault ribbons, facilitating dislocation cross-slip. As shown in Fig. 6, the dislocation density increases with the amount of reduction in thickness. The underlying microstructure evolution can be revealed by the work-hardening curves in Fig. 6(d). The strain at the minimum work-hardening rate, occurring slightly after the onset of the α′ martensite transformation, decreases with the increased dislocation density. We note that the onset of the α′ martensite transformation starts after the initial yielding and the yield stress is determined by self-interaction of dislocations. The dislocation density can be estimated by the Orowan equation[33]:

$$\sigma_y = \sigma_0 + \lambda \mu b \sqrt{\rho} \qquad (5)$$

where $\sigma_y$ and $\sigma_0$ are the yield stresses with or without initial dislocations, $\mu$ the shear modulus, $b$ the Burger's vector, $\rho$ the dislocation density and $\lambda$ the numerical constant in a range of 0.3 to 0.6 for different FCC metals. The estimated dislocation density ranges from ~ $10^{15}$ to ~ $10^{16}$ m$^{-2}$ for the hot-rolled samples with 20% to 40% thickness reduction. The slope of the work-hardening rate in the upturn regime, indicating the rate of phase transformation, is substantially increased by the high density of dislocations. The observations lead to the conclusion that a high density of dislocations inside the affected austenite promotes phase transformation. This may suggest that a pre-stored high density of dislocations cannot suppress martensitic phase transformation by hampering emission of new dislocations from the hetero-interface or blocking motion of the preexisting dislocations.

Based on our *in-situ* TEM study, nucleation of α′ martensite in pristine austenite is always associated with extensive partial dislocation glide, which cannot be readily realized in austenite when a very high dislocation density exists prior to transformation. In contrast, existing dislocations that are located in front of the hetero-interface can be effectively driven by the high misfit elastic stress during transformation, as shown in Fig. 4. The stored dislocations, gliding on a variety of slip systems, can fully accommodate the phase transformation strain as well. On the other hand, the high stress field associated with a high density of dislocations adds to the phase transformation driving force, as suggested by the *in-situ* study in Fig. 5. This agrees with the early theoretical analysis that the stored energy associated with the elastic strain of a dislocation can effectively reduce the martensite nucleation barrier [32]. From both the kinematic and energetic views, high density of pre-stored dislocations can enhance the rate of phase transformation, as implied by the work-hardening curves of hot-rolled samples (Fig. 6(d)).

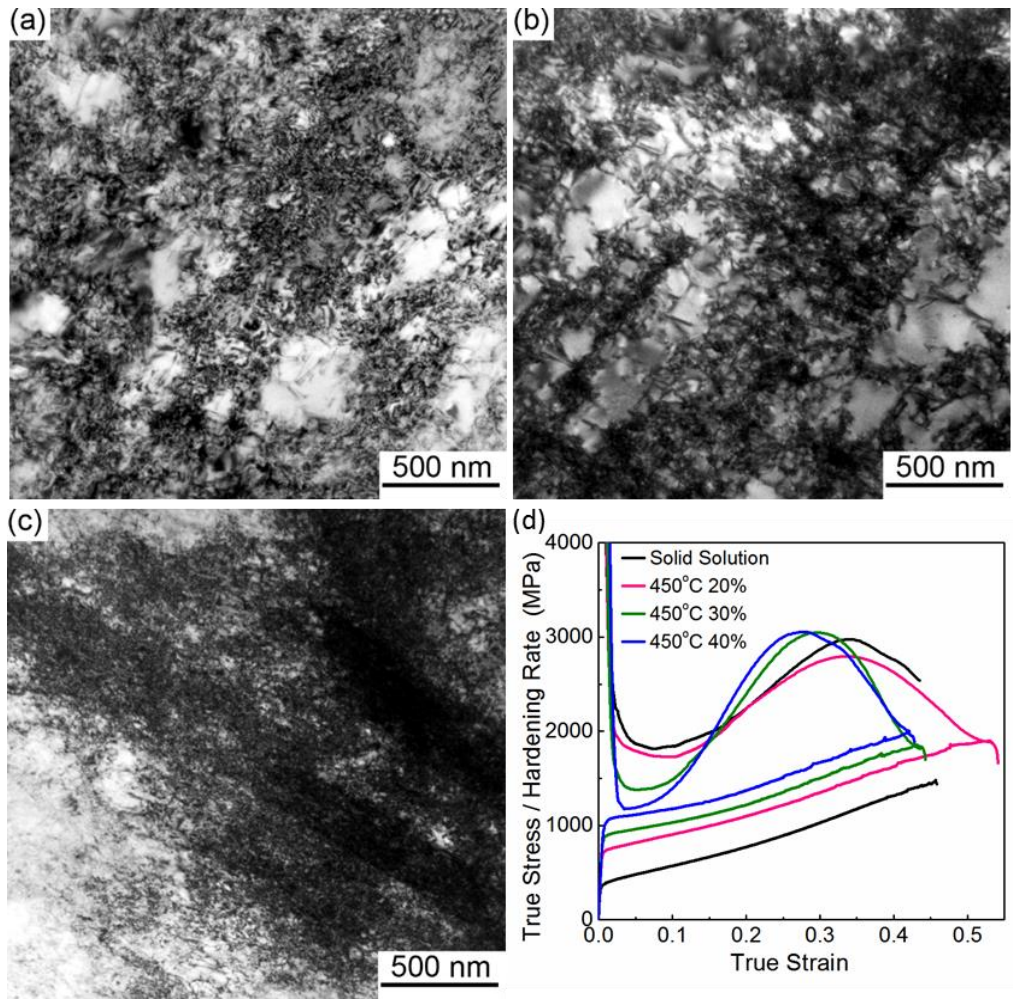

Figure 6. Bright field TEM images of TRIP steels with (a) 20%, (b) 30%, (c) 40% preceding hot rolling thickness reduction at 450 °C. (d) Stress-strain and work-hardening curves of the test samples taken after various thickness reductions.

## 4. Conclusions

We investigated the microscale interactions between dislocations and α′ martensite using in-situ electron microscopy. Local stress concentrations, *e.g.* notches or dislocation pile-ups, render free edges and austenite grain boundaries favorable nucleation sites for α′ martensite. High numbers of dislocations are emitted from the moving martensite-austenite hetero-interfaces during the transformation, resulting in the increase of the dislocation density in the austenite. A kinematic analysis reveals that new dislocations must be emitted from the moving martensite-austenite interface to compensate for the strain misfit associated with the athermal phase transformation. The fact that the phase boundaries move during the transformation leads to a situation where large numbers of new dislocations are continuously pumped into the surrounding austenite matrix without creating local accumulation effects such as observed at immobile interfaces. This mechanism renders the total deformation compatible and hardens the host phase. The permanently moving position of these dislocation sources reduces local damage accumulation due to the permanent shift of the intense slipping and associated plastic relaxation on the softer austenite side of the interfacial region. A well dispersed martensite distribution thus renders plastic deformation more uniform throughout the austenitic matrix, which explains the good combination of strength and ductility of TRIP steels.

**Acknowledgments**

J. Liu acknowledges the support from the financial support from National Natural Science Foundation of China (No. 11572281, 11202183 and u1434202) and the State Key Laboratory for Strength and Vibration of Mechanical Structures Program of Xi'an Jiaotong University (SV2014-KF-13).

**References**

[1] E.R. Parker, V.F. Zackay, Fundamentals of alloy design, CRC Crit. Rev. Solid State Sci. 4(1-4) (1973) 591-613.

[2] E.R. Pakker, V.F. Zackay, Materials science of modern steels, Prog. Solid State Ch. 9(0) (1975) 105-138.

[3] I.Y. Georgieva, Trip steels — a new class of high-strength steels with high plasticity, Met Sci Heat Treat 18(3) (1976) 209-218.

[4] V.F. Zackay, M.D. Bhandarkar, E.R. Parker, The role of deformation-induced phase transformations in the plasticity of some iron-base alloys, in: J. Burke, V. Weiss (Eds.), Advances in Deformation Processing, Springer US1978, pp. 351-404.

[5] A.J. Bogers, W.G. Burgers, Partial dislocations on the {110} planes in the B.C.C. lattice and the transition of the F.C.C. into the B.C.C. lattice, Acta Metall. Mater. 12(2) (1964) 255-261.

[6] G.B. Olson, M. Cohen, A general mechanism of martensitic nucleation: Part II. FCC → BCC and other martensitic transformations, Metall. Trans. A 7(12) (1976) 1905-1914.

[7] S. Chatterjee, H.S. Wang, J.R. Yang, H.K.D.H. Bhadeshia, Mechanical stabilisation of austenite, Mater. Sci. Technol. 22(6) (2006) 641-644.

[8] P.J. Jacques, Transformation-induced plasticity for high strength formable steels, Curr. Opin. Solid State Mater. Sci. 8(3–4) (2004) 259-265.

[9] S.H. Lee, J.C. Lee, J.Y. Choi, W.J. Nam, Effects of deformation strain and aging temperature on strain aging behavior in a 304 stainless steel, Met. Mater. Int. 16(1) (2010) 21-26.

[10] D. Fahr, Stress- and strain-induced formation of martensite and its effects on strength and ductility of metastable austenitic stainless steels, Metal. Trans. 2(7) (1971) 1883-1892.

[11] M.F. Ashby, The deformation of plastically non-homogeneous materials, Philos. Mag. 21(170) (1970) 399-424.

[12] C. Syn, B. Fultz, J. Morris, Mechanical stability of retained austenite in tempered 9Ni steel, Metall. Trans. A 9(11) (1978) 1635-1640.

[13] P. Jacques, F. Delannay, J. Ladrière, On the influence of interactions between phases on the mechanical stability of retained austenite in transformation-induced plasticity multiphase steels, Metall. Mater. Trans. A 32(11) (2001) 2759-2768.

[14] M. Somani, P. Juntunen, L. Karjalainen, R. Misra, A. Kyröläinen, Enhanced mechanical properties through reversion in metastable austenitic stainless steels, Metall. Mater. Trans. A 40(3) (2009) 729-744.

[15] G. Olson, M. Cohen, Kinetics of strain-induced martensitic nucleation, Metall. Trans. A 6(4) (1975) 791-795.

[16] L. Murr, K. Staudhammer, S. Hecker, Effects of strain state and strain rate on deformation-induced transformation in 304 stainless steel: Part II. Microstructural study, Metall. Trans. A 13(4) (1982) 627-635.

[17] P.L. Mangonon, G. Thomas, The martensite phases in 304 stainless steel, Metal. Trans. 1(6) (1970) 1577-1586.

[18] G.B. Olson, M. Cohen, A Mechanism for the Strain-Induced Nucleation of Martensitic Transformation, J. Less-Common Metals 28 (1972).

[19] X.-S. Yang, S. Sun, X.-L. Wu, E. Ma, T.-Y. Zhang, Dissecting the mechanism of martensitic transformation via atomic-scale observations, Sci. Rep. 4 (2014) 6141.

[20] H. Fujita, T. Katayama, In-situ observation of strain-induced γ → ε → α′ and γ → α′ martensitic transformations in Fe–Cr–Ni alloys, Mater. Trans. 33(3) (1992) 243-252.

[21] T. Kikuchi, S. Kajiwara, HVEM in situ observation of isothermal martensitic transformation under applied stress, Trans. Jpn. I. Met. 26(12) (1985) 861-868.

[22] S. Kajiwara, Morphology and crystallography of the isothermal martensite transformation in Fe—Ni—Mn alloys, Philos. Mag. A 43(6) (1981) 1483-1503.

[23] K. Ogawa, S. Kajiwara, High-resolution electron microscopic study on atomic arrangements at growing tips of martensite plates and a nucleating martensite in Fe-Ni-Mn and Fe-Cr-C alloys, Mater. Trans. 48(4) (2007) 860-868.

[24] T. Kakeshita, K. Watanabe, T. Tadaki, K. Shimizu, rsquo, ichi, Two kinds of martensitic transformations in thin Foils of an Fe-32.8mass%Ni alloy, Trans. Jpn. I. Met. 23(9) (1982) 535-543.

[25] C. Tasan, M. Diehl, D. Yan, M. Bechtold, F. Roters, L. Schemmann, C. Zheng, N. Peranio, D. Ponge, M. Koyama, An overview of dual-phase steels: advances in microstructure-oriented processing and micromechanically guided design, Ann. Rev. Mater. Res. (0) (2015).

[26] M. Wang, C.C. Tasan, D. Ponge, A. Kostka, D. Raabe, Smaller is less stable: Size effects on twinning vs. transformation of reverted austenite in TRIP-maraging steels, Acta Mater. 79 (2014) 268-281.

[27] C.C. Tasan, M. Diehl, D. Yan, C. Zambaldi, P. Shanthraj, F. Roters, D. Raabe, Integrated experimental–simulation analysis of stress and strain partitioning in multiphase alloys, Acta Mater. 81 (2014) 386-400.

[28] M. Calcagnotto, Y. Adachi, D. Ponge, D. Raabe, Deformation and fracture mechanisms in fine-and ultrafine-grained ferrite/martensite dual-phase steels and the effect of aging, Acta Mater. 59(2) (2011) 658-670.

[29] K.S. Chan, Y.W. Kim, Influence of microstructure on crack-tip micromechanics and fracture behaviors of a two-phase TiAl alloy, Metall. Trans. A 23(6) (1992) 1663-1677.

[30] S. Sun, S. Sakai, H.G. Suzuki, Effect of Si on the microstructure and mechanical properties of as drawn Cu–15Cr in situ composites, Mat. Sci. Eng. A-Struct. 303(1) (2001) 187-196.

[31] S.-H. Kim, H. Kim, N.J. Kim, Brittle intermetallic compound makes ultrastrong low-density steel with large ductility, Nature 518(7537) (2015) 77-79.

[32] D.A. Porter, K.E. Easterling, M. Sherif, Phase transformation in metals and alloys, CRC Press 2009.

[33] U.F. Kocks, H. Mecking, Physics and phenomenology of strain hardening: the FCC case,